\documentclass{article}
\usepackage{amsmath,amssymb}
\usepackage{algorithm}
\usepackage{algpseudocode}
\usepackage{graphicx}
\usepackage{hyperref}
\usepackage{listings}
\usepackage{xcolor}
\usepackage{float}
\usepackage{svg}
\setsvg{inkscapelatex=false}  
\usepackage[T1]{fontenc}

\makeatletter
\def\thanks#1{\protected@xdef\@thanks{\@thanks
        \protect\footnotetext{#1}}}
\makeatother

\title{Designing Scalable Rate Limiting Systems: Algorithms, Architecture, and Distributed Solutions
}
\author{
  \textbf{Bo Guan}\\
  WynerTech Solutions \\
  billy.guan@wynertech.com
}
\date{}

\begin{document}

\maketitle

\begin{abstract}

Designing a rate limiter that is simultaneously accurate, available, and scalable presents a fundamental challenge in distributed systems, primarily due to the trade-offs between algorithmic precision, availability, consistency, and partition tolerance. This article presents a concrete architecture for a distributed rate limiting system in a production-grade environment. Our design chooses the in-memory cache database, the Redis, along with its Sorted Set data structure, which provides $O(log (N))$ time complexity operation for the key-value pair dataset with efficiency and low latency, and maintains precision. The core contribution is quantifying the accuracy and memory cost trade-off of the chosen Rolling Window as the implemented rate limiting algorithm against the Token Bucket and Fixed Window algorithms. In addition, we explain how server-side Lua scripting is critical to bundling cleanup, counting, and insertion into a single atomic operation, thereby eliminating race conditions in concurrent environments. In the system architecture, we propose a three-layer architecture that manages the storage and updating of the limit rules. Through script load by hashing the rule parameters, rules can be changed without modifying the cached scripts. Furthermore, we analyze the deployment of this architecture on a Redis Cluster, which provides the availability and scalability by data sharding and replication. We explain the acceptance of AP (Availability and Partition Tolerance) from the CAP theorem as the pragmatic engineering trade-off for this use case. 

\end{abstract}


\section{Introduction}

Rate limiting is an essential pattern for regulating network traffic in distributed systems, serving as a critical control mechanism for security, economics, and performance \cite{raghavan2007cloud, stanojevic2009load}. Designing a rate limiter that is simultaneously accurate, available, and scalable presents a fundamental challenge in distributed systems \cite{hu2025distributed, he2021scalable}. The problem is not merely one of counting requests but of architecting a stateful coordination service under the constraints of the CAP theorem, where the need for high availability conflicts with the guarantees of strong consistency in a partitioned environment \cite{li2023noah}. While algorithms like Token Bucket and Fixed Window are well-understood, their application in distributed, high-concurrency production systems introduces critical, under-discussed engineering issues: race conditions during concurrent writes, the memory and accuracy trade-off \cite{chen2025scalatap}, and the availability and consistency dilemma when state is shared across nodes.

This article presents a concrete architecture for a distributed production-grade rate limiting service. The core of our contribution is: First, we explicitly analyze the deployment of this architecture on a Redis Cluster and argue that favoring Availability and Partition Tolerance (AP) over strong Consistency is a pragmatic engineering trade-off for this use case. Second, we quantify the memory usage on each rate limiting algorithm and favor the Rolling Window which provides precision at a low memory cost. Third, we detail how server-side Lua scripting is essential to bundling cleanup, counting, and insertion into a single atomic operation, thereby eliminating race conditions. Our design centers on the Rolling Window algorithm implemented with Redis Sorted Sets by the Lua scripts, which occupies low memory on the server-side (8 bytes per request) and provides $O(log(N))$ operations for efficient state management. 

\begin{figure}[H]
    \centering
    \includegraphics[width=1.2\textwidth]{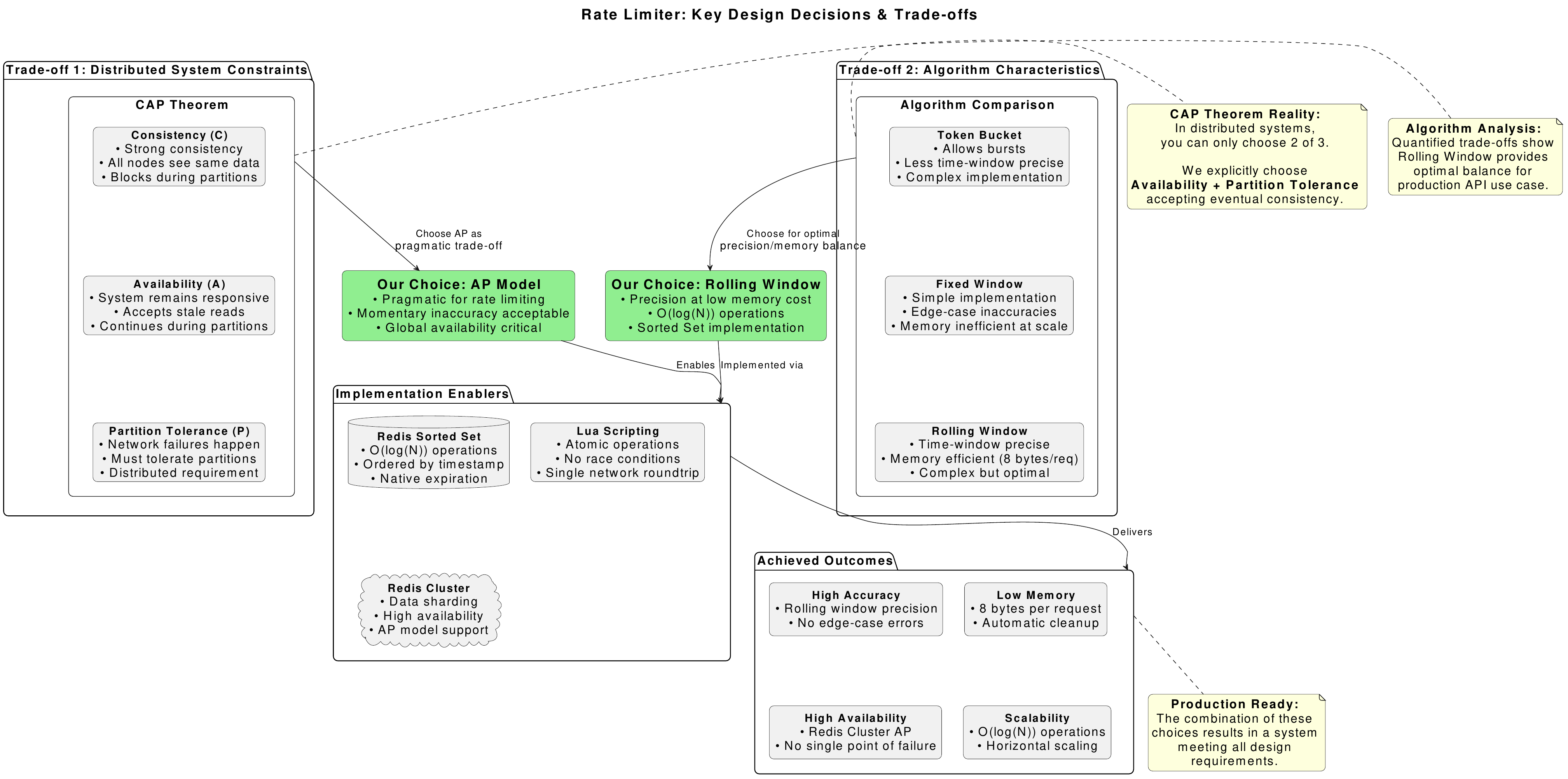}
    \caption{Design decisions and trade-offs in the distributed rate limiter architecture}
    \label{fig:design-tradeoffs}
\end{figure}

Designing a production-grade rate limiter involves navigating fundamental trade-offs in distributed systems. As shown in Figure \ref{fig:design-tradeoffs}, our design makes explicit choices for two primary trade-off spaces: (1) the CAP theorem constraints for distributed systems, where we explicitly choose Availability and Partition Tolerance (AP) over strong Consistency as a pragmatic engineering decision for rate limiting (explained in Section \ref{redis_cluster}); and (2) the algorithmic trade-off space, where we select the Rolling Window algorithm for its optimal balance of time-window precision and memory efficiency compared to Token Bucket and Fixed Window alternatives (explained in Section \ref{freq_limiter}). These decisions are implemented through Redis Sorted Sets with Lua scripting for atomic operations and Redis Cluster for the AP deployment model, achieving the target outcomes of high accuracy, low memory usage, high availability, and scalability.

However, rate limiters are not a universal solution. As noted by Stripe's engineering team, rate limiters are appropriate when users can afford to change the pace at which they hit your API endpoints without affecting the outcome \cite{Tarjan2017Scaling_blog}. This makes them ideal for APIs where requests can be retried or delayed. In contrast, for real-time events where request pacing is impossible (e.g., live notifications, financial ticks, or gaming actions), a simple rate limiter is insufficient—these scenarios often require scaling infrastructure capacity or implementing specialized queuing and load-shedding strategies. This article focuses on the design of scalable rate limiters for the former, the high-volume API use case. It functions by enforcing a predefined limit on the number of requests a user, service, or IP address can make within a specific time window. It protects backend resources from being overwhelmed by traffic spikes or malicious attacks like Denial of Service (DoS), ensuring resource availability. This control is not merely a technical safeguard, but a core business imperative. From a financial perspective, it directly manages operational costs by preventing excessive use of paid third-party APIs and cloud services. Furthermore, it guarantees a fair and consistent quality of service for all users by preventing any single actor from monopolizing the system capacity and degrading the experience for others. In summary, the rate limiter has the following utilities:

\begin{itemize}
\item \textbf{Security:} Protects against resource starvation and Denial of Service (DoS) attacks.
\item \textbf{Cost Control:} Manages expenses from external APIs and infrastructure.
\item \textbf{Fairness:} Ensures an equitable distribution of resource and service quality.
\end{itemize}

Building a robust rate limiter for a high-scale environment introduces a complex set of engineering trade-offs. The design must simultaneously address the challenges of performance, reliability, and scalability under a heavy, continuous write load. The core requirement is to make accurate throttle decisions with extremely low latency to avoid adding significant overhead to HTTP response times. This must be achieved while maintaining memory efficiency and state consistency across a distributed cluster of servers, a non-trivial problem in a multi-threaded, multi-process environment. The system itself must be highly available, resilient to partial failures, and flexible enough to support a variety of throttling rules—such as by IP, user ID, or API endpoint, and adapt to rule change—without structural changes. Finally, it must provide clear communication to the client, returning appropriate error messages when a request is denied. In summary, the design requirements for the distributed rate limiter are as follows:

\begin{itemize}
\item \textbf{Good accuracy:} Accurate throttle decisions.
\item \textbf{Low latency and good memory efficiency:} Minimal impact on response time to requests.
\item \textbf{Distributed coordination with concurrency handling:} Concurrency handling across multiple processes for requests.
\item \textbf{High availability and flexibility:} Resilient to failures and adaptable to various and varying throttle rules.
\item \textbf{Clear client communication:} Explicit feedback for throttled requests.
\end{itemize}

While Hayes et al.~\cite{Hayes2015Better} demonstrated Redis Sorted Sets for rate limiting, this article extends that foundation with production-oriented contributions: quantitative memory-accuracy trade-off analysis, a three-layer dynamic rule management architecture, and explicit CAP theorem trade-off justification for AP-model deployment in distributed environments.

We structure the article as follows. Section 2 provides the necessary background on rate-limiting algorithms and their trade-offs. Section 3 presents our system architecture and the detailed implementation of Redis Sorted Set with Lua scripts. Section 4 analyzes distributed challenges, including our rationale for the AP model and scaling with Redis Cluster. Section 5 discusses performance considerations. We conclude in Section 6.

\section{Algorithmic Foundations}\label{freq_limiter}
In this section, we introduce the foundational algorithms for implementing the rate limiting. We introduce different rate limiter methodologies: token bucket algorithm, fixed window counter, and rolling window counter. In addition, in Section \ref{con_limiter}, we explain the concurrent request limiter to overcome the situation where there are multiple concurrent request processes. We explain each algorithm's principle and their use scenarios.

\subsection{Token Bucket Algorithm}
The Token Bucket algorithm has been widely applied for rate limiting. There are buckets filled with tokens for each user. The number of tokens for a user's bucket is initialized by \textit{capacity} at the beginning; every time the user makes a request action, the tokens will be refilled by the refilling \textit{rate} multiplied by the time since the stored timestamp \textit{last\_refill}, but must always be smaller or equal to the \textit{capacity}. The user's request will be processed if there are enough tokens in the bucket and one token (or more tokens for a more expensive action) will be taken for the request. If there are not enough tokens for the request, the request will be discarded. The algorithm is summarized in Algorithm \ref{alg_1} below.
\begin{algorithm}[H]
\caption{Token Bucket Rate Limiting (In-Memory)}
\begin{algorithmic}[1]
\State Initialize $capacity \gets capacity$, $rate \gets rate$, $user\_id \gets user\_id$
\State $bucket\_key \gets \text{``token\_bucket:''} + user\_id$
\State $buckets \gets \{\}$ \Comment{Global dictionary: key $\to$ (tokens, last\_refill)}

\Procedure{allowRequest}{$tokens\_required$}
    \If{$bucket\_key \notin buckets$}
        \State $buckets[bucket\_key] \gets (capacity, now())$
    \EndIf
    
    \State $(tokens, last\_refill) \gets buckets[bucket\_key]$
    \State $current\_time \gets now()$
    \State $time\_passed \gets current\_time - last\_refill$
    \State $refill\_tokens \gets time\_passed \times rate$
    \State $tokens \gets \min(capacity, tokens + refill\_tokens)$
    \State $last\_refill \gets current\_time$
    
    \If{$tokens \geq tokens\_required$}
        \State $tokens \gets tokens - tokens\_required$
        \State $buckets[bucket\_key] \gets (tokens, last\_refill)$
        \State \Return true
    \Else
        \State $buckets[bucket\_key] \gets (tokens, last\_refill)$
        \State \Return false
    \EndIf
\EndProcedure
\end{algorithmic}
\label{alg_1}
\end{algorithm}

The token bucket algorithm is memory efficient and suitable for handling burst traffic, where user requests can be allowed only if there are enough tokens left in the bucket. The algorithm has two tuning parameters $(capacity, rate)$, which is easy to implement. For example, Amazon API gateway throttles requests to the API server using the token bucket algorithm \cite{AWS2025Throttle}. Strip also uses the token bucket algorithm for rate limiting \cite{Tarjan2017Scaling_blog}. 

\subsection{Fixed Window Counter}
The fixed window counter is the algorithm that counts the frequency of user's requests in each fixed-sized time window. The algorithm is efficient in counting each user's request; however, it fails to handle burst requests at window boundaries, where the total number of requests in two time windows near the boundary could exceed the maximum allowed.
\begin{algorithm}[H]
\caption{Fixed Window Counter (In-Memory)}
\begin{algorithmic}[1]
\State Initialize $window\_size \gets window\_size$, $max\_requests \gets max\_requests$, $user\_id \gets user\_id$
\State $window\_key \gets \text{``fixed\_window:''} + user\_id$
\State $windows \gets \{\}$ \Comment{Global dictionary: key $\to$ (count, window\_start)}

\Procedure{allowRequest}{}
    \State $current\_time \gets now()$
    \State $current\_window \gets \lfloor current\_time / window\_size \rfloor$
    \State $window\_start \gets current\_window \times window\_size$
    
    \If{$window\_key \notin windows$}
        \State $windows[window\_key] \gets (0, window\_start)$
    \EndIf
    
    \State $(count, stored\_window) \gets windows[window\_key]$
    
    \If{$stored\_window < window\_start$}
        \State \Comment{New time window, reset counter}
        \State $count \gets 0$
        \State $stored\_window \gets window\_start$
    \EndIf
    
    \If{$count < max\_requests$}
        \State $count \gets count + 1$
        \State $windows[window\_key] \gets (count, stored\_window)$
        \State \Return true
    \Else
        \State \Return false
    \EndIf
\EndProcedure
\end{algorithmic}
\label{alg_2}
\end{algorithm}

\subsection{Rolling Window Counter}\label{roll_window}

The rolling window counter algorithm overcomes the fixed window counter's vulnerability to request bursts at window boundaries by maintaining a continuously sliding time window. Introduced by Hayes \cite{Hayes2015Better}, this approach tracks individual request timestamps, enabling precise counting of requests within any $T$-second interval rather than fixed calendar-aligned windows. By dynamically evaluating only requests that occurred within the last $T$ seconds, the algorithm prevents evasion of the boundary-based rate limit while providing accurate time-window enforcement, as formalized in Algorithm \ref{alg_3}.
\begin{algorithm}[H]
\caption{Rolling Window Counter (In-Memory)}
\begin{algorithmic}[1]
\State Initialize $window\_size \gets window\_size$, $max\_requests \gets max\_requests$, $user\_id \gets user\_id$
\State $log\_key \gets \text{``rolling\_window''} + user\_id$
\State $request\_logs \gets \{\}$ \Comment{Global dictionary: key $\to$ list of timestamps}

\Procedure{allowRequest}{}
    \State $current\_time \gets now()$
    \State $window\_start \gets current\_time - window\_size$
    
    \If{$log\_key \notin request\_logs$}
        \State $request\_logs[log\_key] \gets [\,]$
    \EndIf
    
    \State $timestamps \gets request\_logs[log\_key]$
    
    \State \Comment{Remove timestamps outside current window}
    \State $i \gets 0$
    \While{$i < \text{len}(timestamps)$ \textbf{and} $timestamps[i] < window\_start$}
        \State $i \gets i + 1$
    \EndWhile
    \State $timestamps \gets timestamps[i:]$
    
    \If{$\text{len}(timestamps) < max\_requests$}
        \State $timestamps.\text{append}(current\_time)$
        \State $request\_logs[log\_key] \gets timestamps$
        \State \Return true
    \Else
        \State \Return false
    \EndIf
\EndProcedure
\end{algorithmic}
\label{alg_3}
\end{algorithm}

Timestamp data can be stored in-memory cache for performance optimization, e.g., Redis (REmote DIctionary Server), which is an in-memory NoSQL key/value caching data store, open source and physically closer to the application \cite{IBM2025What}. Specifically, we can use the data structure, Redis Sorted Set \cite{Hayes2015Better}, where the data is sorted according to the \texttt{SCORE}, e.g., timestamp. And when a new request approaches, it can remove the outdated timestamp data , e.g., previous requests outside the current window, by the command \texttt{ZREMRANGEBYSCORE}. At the end of the request handling, it also sets an automatic timeout on key by the command \texttt{EXPIRE} in case there is no initiating requests. Therefore, the Sorted Set provides memory efficiency. For distributed environments, Redis provides atomic implementation \cite{Hayes2015Better} for race condition prevention, the concurrent requests\footnote{Here the ``concurrent'' means simultaneous requests from different users. Handling concurrent requests from the same user (at the exact same time, e.g., in milliseconds, the system's resolution) will be discussed in Section \ref{con_limiter}, in which we can generate a new random id for each process, e.g., simply replacing the third argument ``now'' by an id number at code line 19 in Algorithm \ref{alg_4} by the generated id, as a member of the concurrent requests, similar to the Multiset (sorted) data structure in C++ language.}, by Algorithm \ref{alg_4}. More details about concurrent handling will be discussed in Section \ref{sec_race_cond}.

In addition, we compare the last request to the current timestamp. If they are too close, we also do not allow the action. For retrieving the last request's timestamp, at line $15$, we use the \texttt{ZREVRANGE} command, which returns a list of members (user id here) in the specified range, optionally with their scores for sorting (timestamp here) if \texttt{WITHSCORE} was used \cite{Redis2025Zrevrange}. The time complexity for the operations is: $O(log(N)+M)$ with $N$ being the number of elements in the Sorted Set and $M$ the number of elements returned. And the time complexity to add the request as a new member of the Sorted Set \texttt{ZADD} is $O(log(N))$. Using the Redis data store can also handle the scale needs of distributed systems, where data will be accessed from multiple users or processes, which will be covered in Section \ref{distr_sys}. 

\begin{algorithm}[H]
\caption{Rolling Window with Redis Sorted Sets}
\begin{algorithmic}[1]
\State Initialize $window\_size \gets window\_size$ \Comment{Rolling window size (seconds)}
\State Initialize $TTL \gets TTL$ \Comment{For removing individual expired requests when time out (seconds)}
\State Initialize $max\_requests \gets max\_requests$ \Comment{Max requests per window}
\State Initialize $min\_interval \gets min\_interval$ \Comment{Minimum interval of any two valid requests (seconds)}
\State $redis\_key \gets \text{``rate\_limiter:''} + user\_id$

\Procedure{allowRequest}{}
    \State $now \gets now()$
    \State $window\_start \gets now - window\_size$
    
    \State \Comment{Atomic Redis operations}
    \State $redis.ZREMRANGEBYSCORE(redis\_key, -\infty, window\_start)$ \Comment{Active cleanup of members outside the rolling window for accurate counting}
    \State $current\_count \gets redis.ZCARD(redis\_key)$
    
    \If{$current\_count \geq max\_requests$}
        \State \Return false \Comment{Too many requests in window}
    \EndIf
    
    \State $last\_request \gets redis.ZREVRANGE(redis\_key, 0, 0, \text{WITHSCORES})$
    \If{$last\_request$ exists and $(now - last\_request[1]) < min\_interval$}
        \State \Return false \Comment{Requests too close together}
    \EndIf
    
    \State $redis.ZADD(redis\_key, now, now)$
    \State $redis.EXPIRE(redis\_key, TTL)$ \Comment{Set a timeout on key}
    \State \Return true
\EndProcedure
\end{algorithmic}
\label{alg_4}
\end{algorithm}

In summary, we choose the in-memory Redis as the data store for storing data rather than using a disk-based storage engine because:
\begin{itemize}
    \item In-memory databases are faster because they can avoid the overhead of encoding in-memory data structures in operating system caches in a form that can be written to disk and managing operating system caches \cite{harizopoulos2018oltp}\footnote{In-memory databases improve the performance is not because they don't need to read from disk. A disk-based storage engine may never need to read from disk if the operating system has enough memory and it caches recently used disk blocks in memory anyway \cite{kleppmann2017designing}.}.
    \item Redis provides a Sorted Set data structure that is further optimized for simple implementation but is difficult to implement with disk-based indexes \cite{kleppmann2017designing}.
\end{itemize}

\begin{figure}[H]
    \centering
    \includegraphics[width=\textwidth]{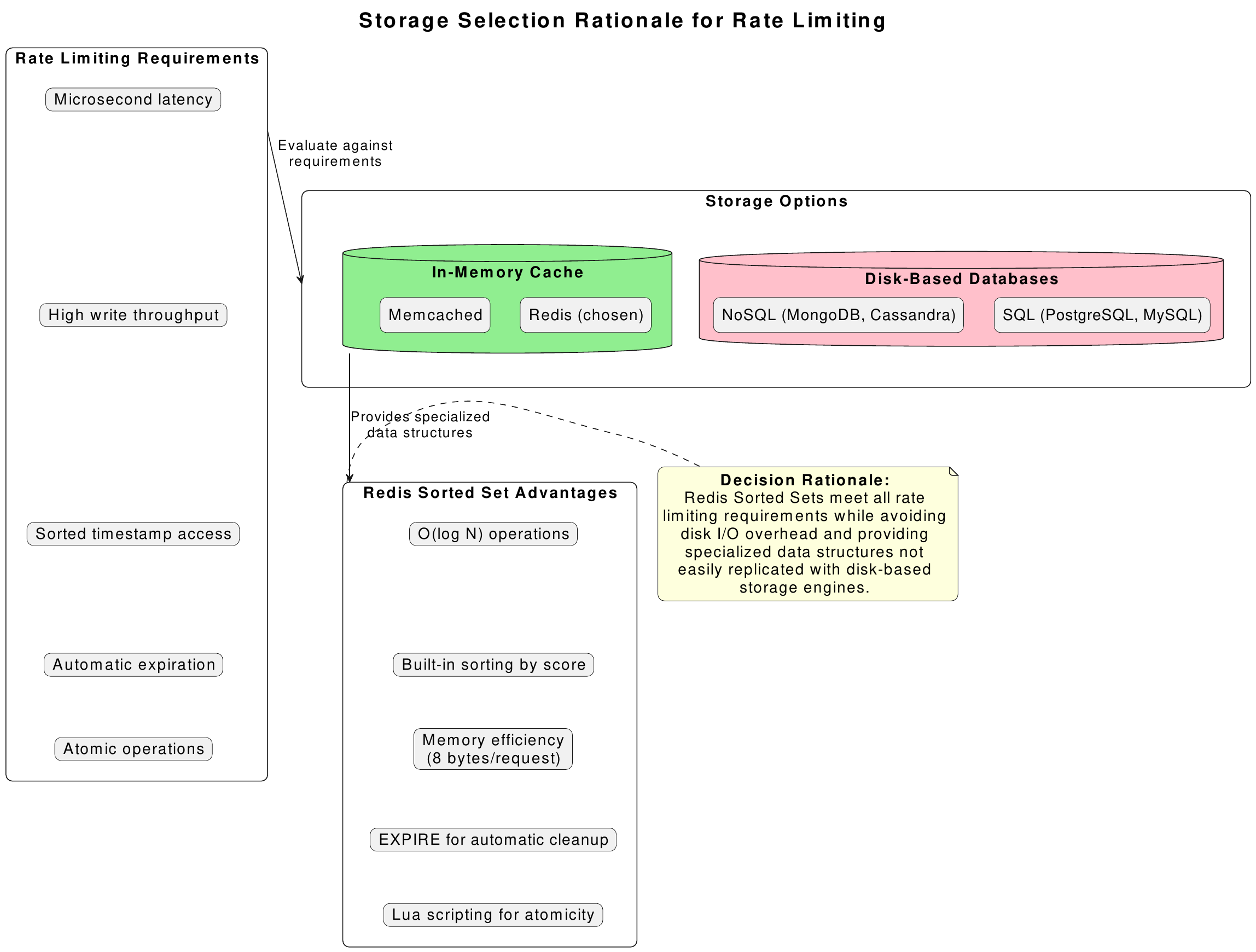}
    \caption{Storage selection rationale for rate limiting implementation. The diagram illustrates the evaluation of storage options against rate limiting requirements, leading to the selection of Redis Sorted Sets for their O(log N) operations, built-in timestamp sorting, memory efficiency, and atomic execution capabilities via Lua scripting—features not easily replicated with disk-based databases.}
    \label{fig:storage-selection}
\end{figure}

\subsection{Comparative Analysis}
The token bucket algorithm allows requests as long as tokens are available and refills tokens at a fixed rate, enabling configurable, predictable burst tolerance—a desirable feature for handling natural user behavior. The fixed window counter algorithm counts requests per time window but suffers from edge burst problems, allowing double the limit when requests span window boundaries. The rolling window algorithm addresses this with precise timestamp tracking, but requires more memory, storing individual timestamps for each request rather than a single counter, and adds cleanup overhead. Concurrent limiters track active requests rather than request rates, enforcing a maximum number of simultaneous operations to protect resources from exhaustion—ideal for scenarios like database connections, file uploads, or long-running API calls where parallel execution must be controlled. The details of the concurrent limiter will be discussed in Section \ref{con_limiter}.

The token bucket algorithm is memory-efficient, storing exactly two double-precision floating-point numbers—tokens and last refill time—requiring 16 bytes per user with no active cleanup as tokens automatically refill. Fixed window counters require 16 bytes per user (an 8-byte counter and an 8-byte window start timestamp, both doubles here\footnote{we assume the storage of counter and timestamp as 8-byte doubles; Actual memory usage may vary slightly due to its internal structures.}) but need periodic window expiration. Rolling window algorithms store individual 8-byte double timestamps for each request, consuming exactly $8L$ bytes per user where $L$ is the request limit, and require active cleanup of expired timestamps. Concurrent limiters store both a timestamp and a request identifier for each active request, consuming $16C$ bytes per user, where $C$ is the concurrent limit, utilizing TTL-based automatic expiration. For one million users with typical limits ($L=100$, $C=50$), rolling windows require 800 MB, token buckets 16 MB, fixed windows 16 MB, and concurrent limiters 800 MB
. Although the rolling window algorithm and the concurrent limiter (which is rolling window based) require more memory than the token bucket or fixed window approaches, this trade-off enables precise time-window enforcement—a critical requirement for API rate limiting where boundary burst protection is essential.

\begin{table}[H]
\centering
\begin{tabular}{|l|cccc|}
\hline
\textbf{Algorithm} & \textbf{Accuracy} & \textbf{Memory} & \textbf{Burst Handling} & \textbf{Primary Use Case} \\
\hline
Token Bucket & Medium & Low & Excellent & API rate limiting \\
Fixed Window & Low & Low & Poor & Simple throttling \\
Rolling Window & High & High & Good & Precise rate enforcement \\
Concurrent Limiter & High & Medium & Excellent & Resource protection \\
\hline
\end{tabular}
\caption{Algorithm Comparison by Characteristics.}
\end{table}

\begin{table}[H]
\centering
\begin{tabular}{|l|cc|}
\hline
\textbf{Algorithm} & \textbf{Per-User Memory} & \textbf{Total (1M Users)} \\
\hline
Token Bucket & 16 bytes & 16 MB \\
Fixed Window & 16 bytes & 16 MB \\
Rolling Window & $8L$ bytes & 800 MB \\
Concurrent Limiter & $16C$ bytes & 800 MB \\
\hline
\end{tabular}
\caption{Memory with Redis Storage Overhead (for rolling window and concurrent limiter). 1M users with typical limits ($L=100$, $C=50$).}
\label{tab:memory-redis}
\end{table}

This article implements the rolling window counter algorithm using Redis Sorted Sets, which provides precise time-window enforcement while addressing distributed system challenges. The implementation resolves concurrency issues through atomic Lua scripting, balances availability-consistency trade-offs via Redis Cluster's AP model, and scales horizontally—all detailed in Section \ref{sys_arch} (System Architecture) and Section \ref{distr_sys} (Distributed Systems Analysis).

\section{System Architecture} \label{sys_arch}

\begin{figure}[H]
    \centering
    \includegraphics[width=1.2\textwidth]{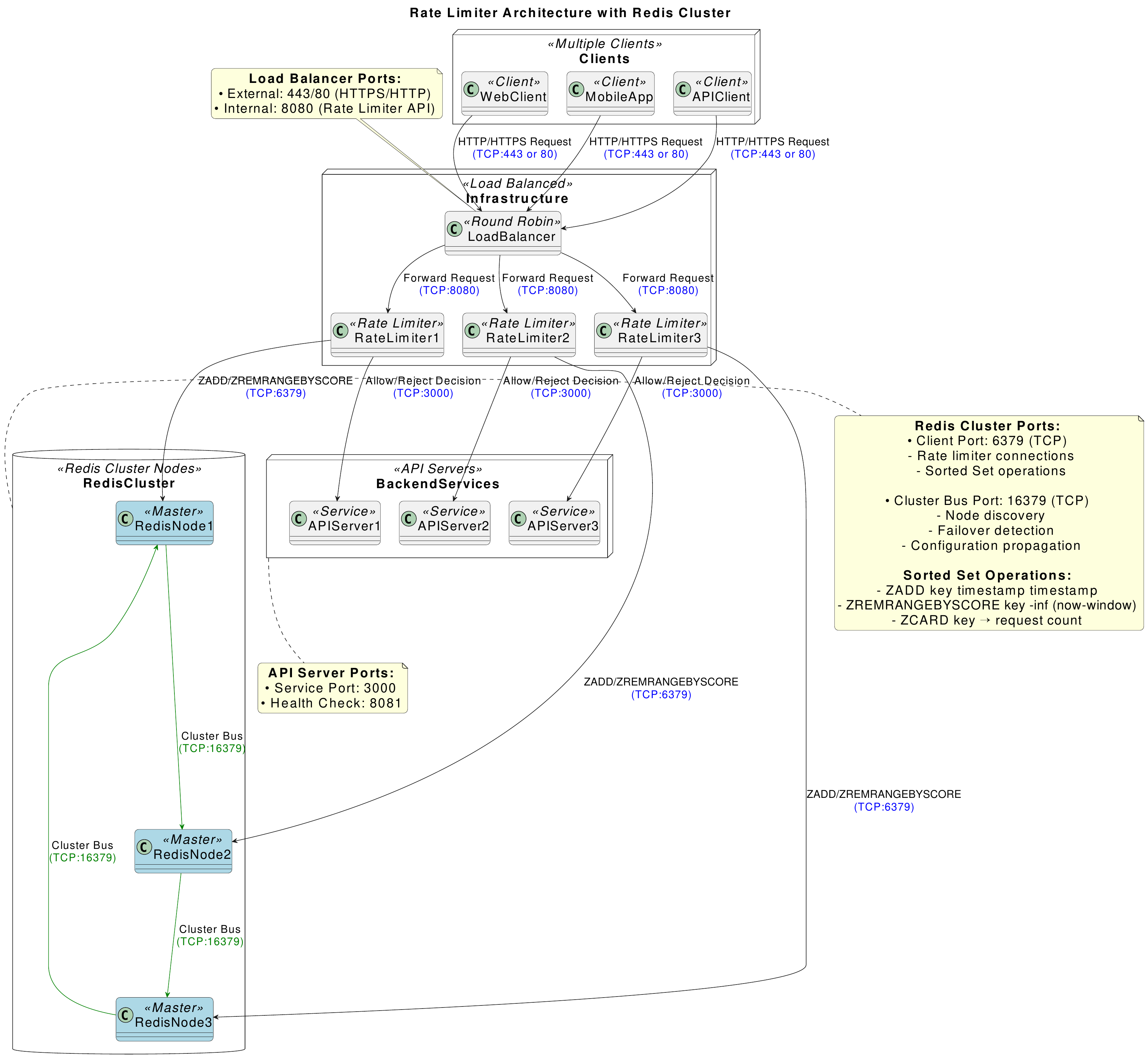}
    \caption{Rate Limiter System Architecture using Redis Sorted Sets for distributed coordination and atomic operations.}
    \label{fig:rate-limiter-arch}
\end{figure}

We design a high availability scale rate limiter system by deploying multiple server nodes, the Redis cluster, and multiple rate limiter instances. As illustrated in Figure~\ref{fig:rate-limiter-arch}, client requests from web, mobile and API sources are load-balanced across multiple rate limiter instances, which coordinate through a Redis cluster (with multiple Redis nodes; more details explained in Section \ref{redis_cluster}) using a Sorted Set data structure and Lua scripts for atomic operations (more details explained in Section \ref{con_limiter}). This design ensures that rate limiting decisions remain availability even when some server node fails by replacing with its replication (more details explained in Section \ref{fault_toler}).

\begin{figure}[H]
    \centering
    \includegraphics[width=1.2\textwidth]{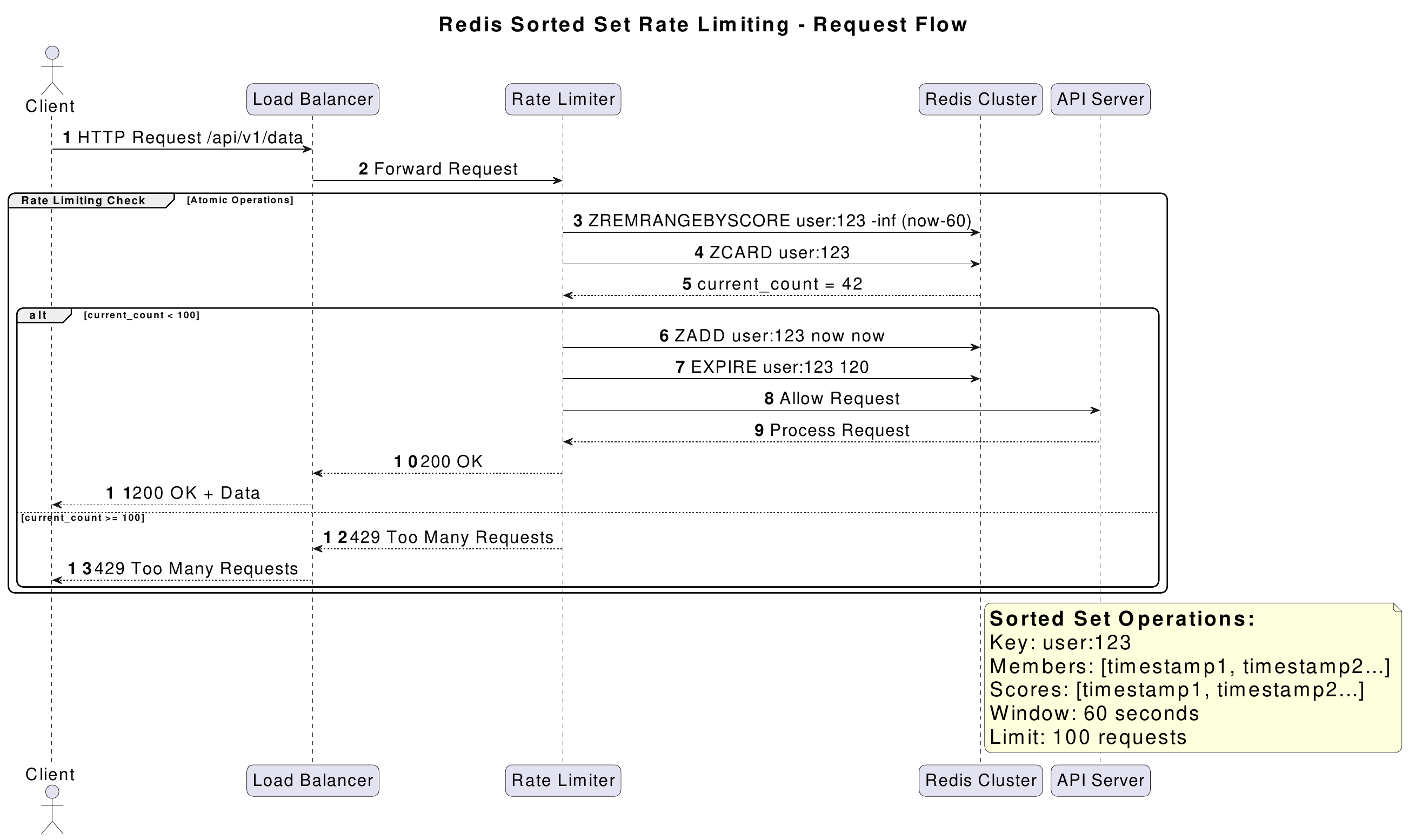}
    \caption{Request Flow Sequence for Redis Sorted Set Rate Limiting demonstrating atomic operations and rolling window enforcement.}
    \label{fig:request-flow-sequence}
\end{figure}

The request flow sequence depicted in Figure~\ref{fig:request-flow-sequence} demonstrates the atomic nature of the rate limiting check: each request triggers a coordinated sequence of Redis operations, including \texttt{ZREMRANGEBYSCORE} to clean outdated entries, \texttt{ZCARD} to count current requests within the rolling window, and \texttt{ZADD} to record new requests—all executed within a single atomic transaction to prevent race conditions, initiated by requests from different users (or processes). Separation of rate limiting middleware from backend API servers allows for independent scaling and fault isolation. This architecture provides true rolling window rate limiting with distributed coordination, with high availability, and race condition prevention. Based on the rolling window counter of Algorithm \ref{alg_4}, it ensures accurate enforcement without the boundary condition vulnerabilities of fixed-window approaches.

\subsection{Rule Management Architecture}

\begin{figure}[H]
    \centering
    \includegraphics[width=1.2\textwidth]{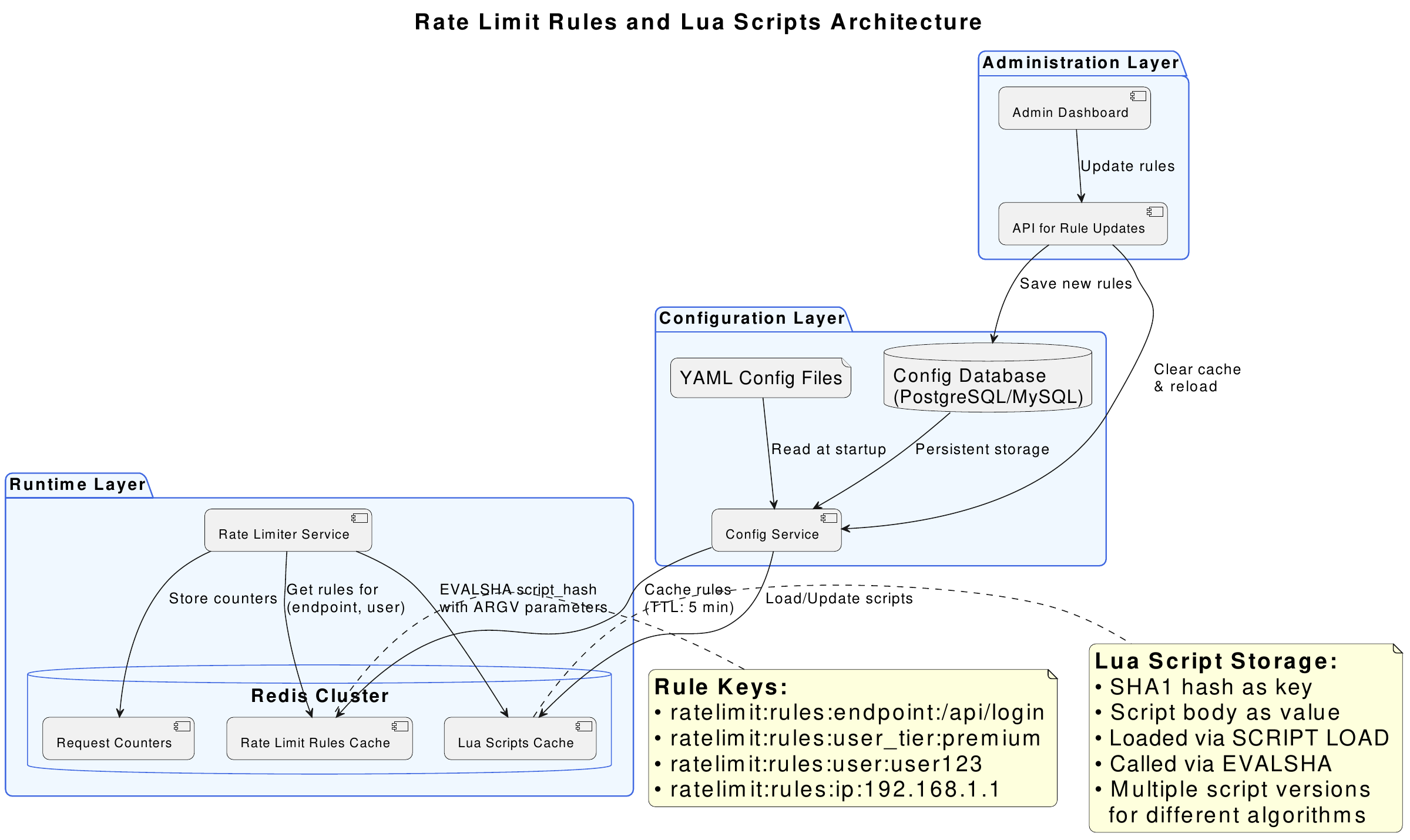}
    \caption{A three-layer rate limit rule management architecture: configuration, runtime enforcement, and administration.}
    \label{fig:rate-limit-rule-manage}
\end{figure}

Following the introduction of the overall system architecture and request flow sequence, a critical question arises: how can rate limit rules be stored or updated dynamically while avoiding script redeployment? The answer lies in a three-layer architecture that separates configuration, runtime enforcement, and administration, as shown in Fig. \ref{fig:rate-limit-rule-manage}. This design enables dynamic rule updates without service disruption while maintaining atomic operations and low latency—essential for production environments where rules must adapt to changing requirements.\footnote{While our algorithms and implementation examples use \texttt{user\_id} for clarity as shown in Section \ref{freq_limiter}, the same patterns apply to any discriminating key: IP addresses, API endpoints, client applications, geographic regions, subscription tiers, or composite keys. This flexibility is achieved through parameterized Lua scripts that accept the key as an argument, enabling different rate limits per dimension without code changes.}

\paragraph{Architectural Layers}
The system implements a clear separation of concerns across three layers:

\begin{itemize}
    \item \textbf{Configuration Layer:} Rate limit rules—defining limits, time windows, and algorithms—are persistently stored in a database \cite{Redis2025Persistence} and cached in Redis with expiration TTL. The Config Service loads these rules into Redis and pre-compiles Lua scripts for each rate limiting algorithm, storing them by \texttt{SHA1} hash for atomic execution \cite{Redis2025Evalsha}.
    
    \item \textbf{Runtime Enforcement Layer:} Rate Limiter services first retrieve applicable rules from cache, then execute the corresponding Lua script via \texttt{EVALSHA}, passing rule parameters as arguments \cite{Redis2025Programmability}. This ensures atomic counter operations on the server-side while maintaining flexibility.
    
    \item \textbf{Administration Layer:} Rule updates occur through a dedicated API and dashboard, where changes are persisted to the database and trigger cache invalidation. This enables dynamic rule changes without service disruption while maintaining performance through Redis caching.
\end{itemize}

\paragraph{Key Design Principles}
The architecture embodies several important design principles that enable its operational flexibility:

\begin{enumerate}
    \item \textbf{Rules as configuration:} Rate limits are defined declaratively and stored separately from executable code, allowing updates without redeployment.
    
    \item \textbf{Lua scripts as parameterized logic:} Scripts serve as templates that receive rule values as \texttt{ARGV} parameters, enabling atomic operations while maintaining separation from rule definitions.
    
    \item \textbf{Separation of concerns:} Rules can change independently of scripts, supporting A/B testing of different algorithms and varying rules per endpoint or user tier.
    
    \item \textbf{Dual caching strategy:} Both rules and scripts are cached in Redis for performance, with scripts stored by \texttt{SHA1} hash for efficient execution via \texttt{EVALSHA}.
    
    \item \textbf{Horizontal scalability:} The stateless nature of rate limiter instances combined with shared Redis state enables consistent behavior across horizontally scaled deployments.
\end{enumerate}

This architectural approach provides several operational benefits: rules can be updated in real-time without code changes, different rate limiting algorithms can be tested simultaneously, and atomicity is preserved through server-side Lua script execution. The configuration patterns are similar to those used in Lyft's open-source rate limiting component \cite{Lyft2020ratelimit}, demonstrating production viability.

\section{Extension for Solving Concurrency and Solutions for Distributed Systems}\label{distr_sys}

\subsection{Race Conditions in Distributed Counting} \label{sec_race_cond}
\begin{figure}[H]
    \centering
    \includegraphics[width=0.8\textwidth]{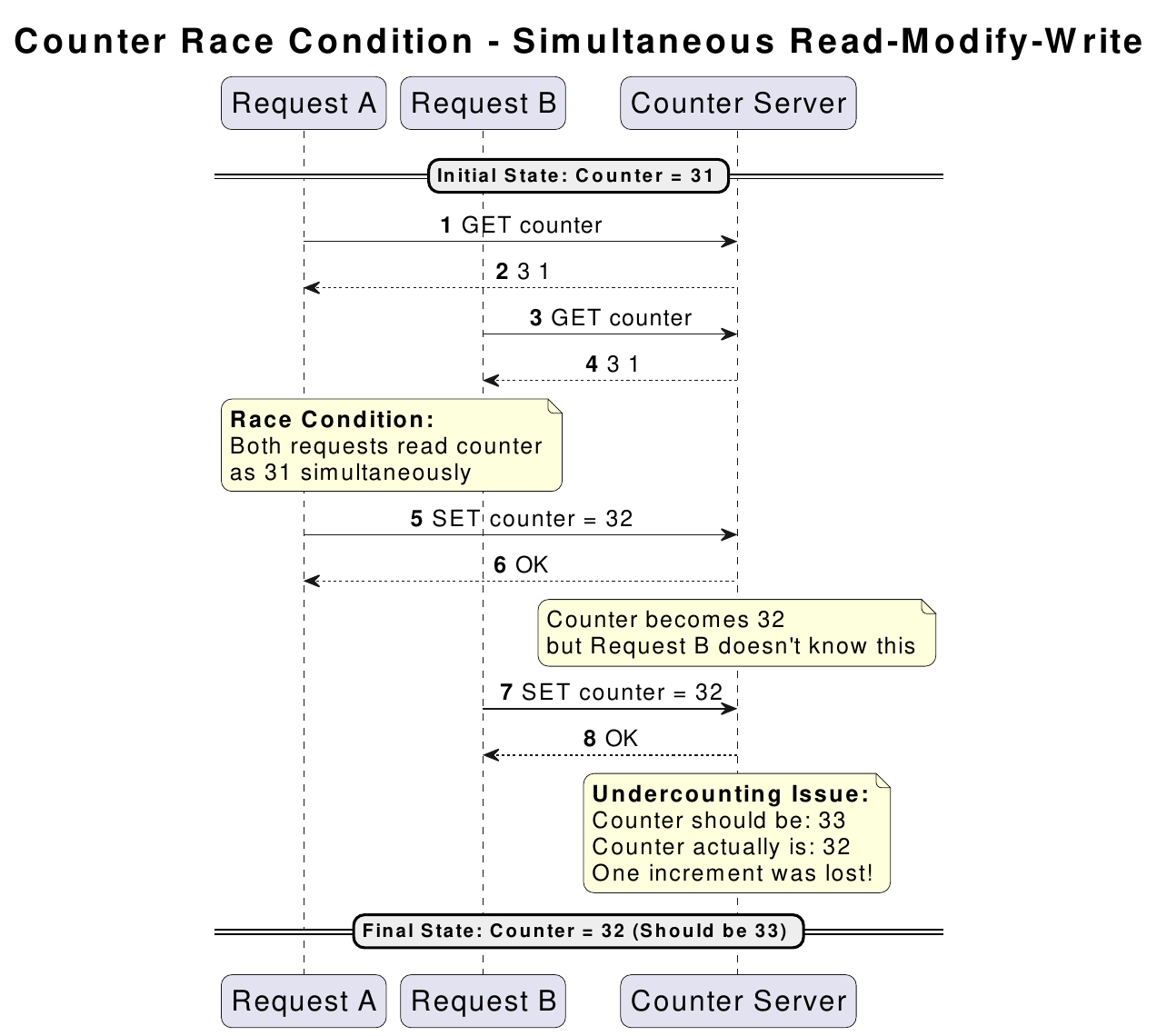}
    \caption{Counter race condition: simultaneous read-modify-write cycles.}
    \label{fig:race-condition}
\end{figure}

We extend the rate limiter design from Section \ref{roll_window}, to solve concurrent requests from the same user, even at the same timestamp (using different processes). Fig. \ref{fig:race-condition} is an example for showing this problem: two requests (either from two different users or the same user from different processes) simultaneously read the counter variable from the server before either writes back for increment, causing undercounting issue: both requests need to read the current counter value, modify the value, and write the new value back. In Fig. \ref{fig:race-condition}, the counter should have increased from 31 to 33, because of the race condition, both read the counter as 31 (but actually one of them is in the process updated to 32) and ``increment'' it to 32.

As shown in the above example, the lost update can occur if an request reads value from the database, modifies it, and writes back the modified value (a \textit{read-modify-write cycle}), e.g., incrementing a counter in this case. Some existing solutions in databases to prevent lost updates due to race conditions are:
\begin{itemize}
    \item Some relational databases provide \textit{atomic} operations, for instance, by increment operations, which is concurrency-safe for the most of time, by putting an exclusive lock on the object when it is read in order that no other transaction can read it until the update has been committed, or simply forcing all atomic operations to be executed on a single thread. However, this solution might sometimes be potentially unsafe and create bugs that are difficult to debug \cite{kleppmann2017designing, Wiger2010An}.
    \item The \textit{compare-and-set} solution prevents two clients from concurrently updating the same variable by allowing a write to occur only if the value has not changed since you last read it. For example, to prevent two requests concurrently updating the same user\_id's counter value, you might check the counter value that has not changed since the server started editing it.
    If the value has changed and no longer matches the ``old value'', this update will have no effect. So, it is your duty to check whether the update took effect and retry if necessary. However, this may or may not be safe depending on the implementation of the server's database \cite{kleppmann2017designing}. 
    \item Some databases have an automatic lost update detection feature that aborts the transaction and forces it to retry its \textit{read-modify-write cycle} if it detects a lost update.
\end{itemize}

Relying on a database's atomic operation and the lost update detection feature might work in some cases. However, it is always the application developer's responsibility to carefully handle the concrete issue with available resources at hand. We are about to talk about the atomic design by implementing the Lua script of Redis server, which is to overcome the concurrency issue and guarantee low latency in rate limiting in the following.

We now apply these atomic operation principles to implement a concurrent request limiter—a distinct rate limiting pattern that controls parallel request concurrency rather than request frequency.


\subsection{Concurrent Request Limiter: Managing Active Connections}\label{con_limiter}

Unlike the frequency-based rate limiters in Section \ref{freq_limiter} (which control requests per time window), the concurrent request limiter controls the \textbf{number of simultaneously active requests} per user \cite{Tarjan2017Scaling_blog}. This prevents resource exhaustion by bounding parallel request concurrency. Stripe's implementation uses the same Redis Sorted Set and Lua scripting pattern as token bucket rate limiting, but with a critical difference: it adds requests to a sorted set upon arrival and removes them upon completion, using \texttt{ZCARD} to enforce a concurrency limit \cite{Tarjan2017Scaling}. The atomic operations provided by Redis Lua scripting, which we introduced in Section \ref{sec_race_cond} to solve race conditions, are equally essential here to ensure accurate counting of concurrent requests.


Thus, to distinguish individual concurrent connections, we require an additional \textbf{request identifier} alongside the \textit{user id}. This allows the system to remove specific requests from the active set when they complete, rather than waiting for $TTL$ expiration.

First, we explain how to count the concurrent requests accurately. Similarly to the basic rolling window counter-based limiter in Algorithm \ref{alg_4}, we also apply the Redis Sorted Set data structure, which has a friendly sorting for members by their \texttt{SCORE}: \textit{user id} as the \texttt{KEY}, timestamp as the \texttt{SCORE} for members' sorting and range operations. We actively remove old requests efficiently outside the window, in the range $(-\infty, \text{current timestamp} - window\_size]$ by using the \texttt{ZREMRANGEBYSCORE} command), to accurately count the requests accumulated in the current window. Here we use the \textit{request id} as the \texttt{MEMBER}, generated by a long enough random id sequence for each request of the user, which guarantees the uniqueness: for instance, a 4 random bytes create $2^{32}$ (over 4.2 billion) possible numbers, and hence the probability of a collision for a single user is astronomically low for the intended use case, e.g., 50 maximum concurrent requests. In the end, we use the \texttt{ZREM} command to set the key's lifetime associated with the user id by $TTL$ seconds. Only when a user becomes completely inactive will their key expire after $TTL$ seconds; otherwise, every successful request extends the user key's lifetime by another $TTL$ seconds. Thus, the algorithm prevents memory leaks by setting the expiration of the user key.
The sketch of the concurrent requests limiter is shown in Algorithm \ref{alg_5}\footnote{Note that in the design, blocked actions still count as actions, so if the number of requests from a user continually exceeds the rate limiter, none of their actions will be allowed until timeout $TTL$ since the last valid request.}

Although the Sorted Set data structure can accurately track the number requests of each user even if it happens concurrently, how could we really solve the race condition brought by the concurrency (where multiple processes are trying to run and increment the counter at the same time)? In order to confront the concurrent requests' race condition for the counter, Redis is able to execute Lua scripts on the server side, which are executed atomically \cite{Redis2025Atomicity}. We can put both read and write operations that need to be performed on the server side, e.g., to increment the counter with minimal latency \cite{Redis2025Pipelining}, by using the \texttt{ZCARD} and \texttt{ZADD} command in the Lua script. As a result, no other script can run while a script is running\footnote{The atomic operation logic, which includes the pseudocode lines 10 to 18 in Algorithm \ref{alg_5}, can be executed in the Lua scripts on the server side \cite{Tarjan2017Scaling, Redis2025Eval}.}.

\begin{algorithm}[H]
\caption{Concurrent Request Limiting with Redis Sorted Sets}
\begin{algorithmic}[1]
\State Initialize $window\_size \gets window\_size$ \Comment{For rate limiting window size (seconds)}
\State Initialize $TTL \gets TTL$ \Comment{For removing individual expired requests when time out (seconds)}
\State Initialize $max\_requests \gets max\_requests$ \Comment{Maximum concurrent requests per user}

\Procedure{checkConcurrentRequest}{$user\_id$}
    \State $current\_time \gets now()$
    \State $request\_id \gets generateUniqueId()$ \Comment{4+ random bytes}
    \State $redis\_key \gets \text{``concurrent\_limiter:"} + user\_id$
    
    \State \Comment{Atomic cleanup and counting}
    \State $redis.ZREMRANGEBYSCORE(redis\_key, -\infty, current\_time - window\_size)$ \Comment{Active cleanup of members outside the rolling window for accurate counting}
    \State $current\_count \gets redis.ZCARD(redis\_key)$
    \State $allowed \gets current\_count < max\_requests$
    
    \If{$allowed$}
        \State $redis.ZADD(redis\_key, current\_time, request\_id)$
        \State $redis.EXPIRE(redis\_key, TTL)$ \Comment{Passive cleanup of abandoned keys and prevent memory leaks}
        \State \Return $(allowed, request\_id)$
    \Else
        \State \Return $(allowed, null)$
    \EndIf
\EndProcedure

\Procedure{completeRequest}{$user\_id, request\_id$}
    \If{$request\_id$ is not null}
        \State $redis\_key \gets``concurrent\_limiter:" + user\_id$
        \State $redis.ZREM(redis\_key, request\_id)$ \Comment{Remove completed request}
    \EndIf
\EndProcedure

\Procedure{doRequest}{$user\_id$, $businessLogic$}
    \State $(allowed, req\_id) \gets \Call{checkConcurrentRequest}{user\_id}$
    \If{not $allowed$} 
        \State \textbf{raise} RateLimitError 
    \EndIf
    \State $businessLogic()$
    \State \Call{completeRequest}{$user\_id$, $req\_id$}
\EndProcedure
\end{algorithmic}
\label{alg_5}
\end{algorithm}


\subsection{Availability and Consistency Trade-off: Redis Cluster}\label{redis_cluster}

Distributed methods can increase system availability and scalability. Data sharding and replication techniques are commonly used together. The \textit{Redis Cluster} is a distributed version of Redis that provides a way to run a Redis installation where data is automatically sharded and replicated across multiple Redis nodes \cite{Redis2025Scale}, enabling horizontal scaling and improving performance. Redis Cluster provides read/write scalability by using \textit{hash slot} (map keys to hash slots by taking the hash value's modulo), but not \textit{consistent hashing}. This makes it easy to add and remove cluster nodes by adjusting the pre-caculated hash slots assignment (e.g., equally assignment) for each node determined by the administrator of the Redis Cluster. However, \textit{consistent hashing}, which distributes the hash value to the first node in the clockwise direction on the hash ring, can cause data skew, leading to an imbalance distribution of keys when one node fails, e.g., due to crash or network partition \cite{Namuag2025Hash}. Redis Cluster uses a \textit{leader-replica replication} model where every hash slot has the leader and its replicas to maintain availability: if the leader node fails, the replica node will be promoted as the new leader and the system will continue to operate correctly. There is an exception for the computation of the hash slot, the hash tags, which are a way to ensure that multiple keys are allocated in the same hash slot. This is used to implement multi-key operations in Redis Cluster \cite{Redis2025Scale, Redis2025Redis_cluster}.

Thus, multiple rate limiter instances (which are stateless) can share a Redis cluster, which can provide:
\begin{itemize}
    \item High \textit{scalability} through data sharding and replication.
    \item Some degree of \textit{consistency} rate limiting across all instances by the \textit{replication} between Redis Cluster nodes.
\end{itemize}

Although this configured Redis Cluster method can mitigate the data inconsistency across multiple nodes, high \textit{availability} would sacrifice some degree of \textit{consistency} (or vice versa) when network partitioned by the \textit{CAP theorem} (which states we have to choose between either \textit{consistency} or total \textit{availability} when a network partition occurs \cite{kleppmann2017designing}) or other practical issues, e.g., node failure, and hence Redis Cluster does not guarantee strong \textit{consistency}\footnote{Here the ``\textit{consistency}'' means ``\textit{linearizability}'' in the \textit{CAP theorem}: as soon as one client successfully completes a write, all clients reading from the database must be able to see the value just written \cite{kleppmann2017designing}. Do not confuse different concepts of the \textit{replica consistency} and the \textit{eventual consistency} that arises in asynchronously replicated systems, and the \textit{consistency} in the context of ACID, which refers to a ``good state'' of the database in an application scenario.}. In practical terms, this means that under certain conditions, it is possible that the Redis Cluster loses writes that were acknowledged by the system to the client. 

The first scenario when the Redis Cluster can lose writes is that it uses \textit{asynchronous replication} and the leader sometimes can fail\footnote{For simplification, here we only consider the case of \textit{single-leader replication}, although \textit{multi-leader replication} has advantages on availability and latency over the single-leader mode. Atomic operations can work in a replicated context when a value is concurrently written by different clients by merging together the updates and getting the same results, which especially works well for commutative operations, e.g., incrementing a counter in this case, and hence prevents lost updates across replicas \cite{Riak2014Distributed}, as shown in Fig. \ref{fig:rate-limiter-arch}.\\

For the \textit{single-leader replication}, if the database is partitioned, different partitions may have their leaders located on different nodes, but each partition must have one leader node.}. The leader node first acknowledges the write from the client, but then crashes before sending the write to its replicas; one of the replica nodes that did not receive the write can be promoted to leader later. Thus, the system will lose the write forever since the old leader's unreplicated writes were discarded.

\begin{figure}[H]
    \centering
    \includegraphics[width=1.2\textwidth]{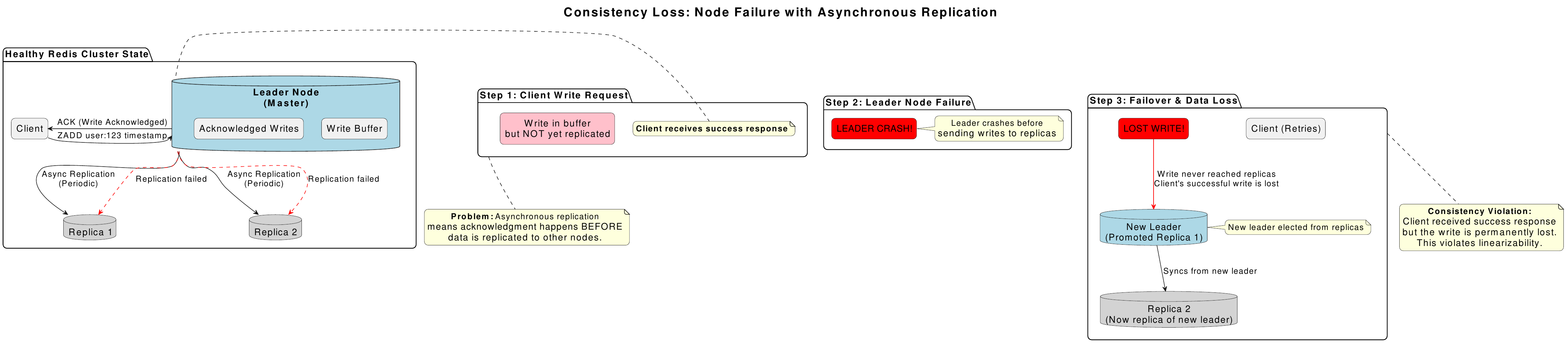}
    \caption{\textbf{Write loss scenario due to leader failure with asynchronous replication in Redis Cluster.} The leader acknowledges a client write but crashes before replicating it to followers. When a replica is promoted to new leader during failover, the client's acknowledged write is permanently lost, violating linearizability. This illustrates one scenario where Redis Cluster (an AP system) may sacrifice consistency for availability.}
    \label{fig:node-failure}
\end{figure}

Other fault scenarios include the \textit{split brain}, which occurs when two nodes both believe that they are the leader. Since if both leaders accept writes and there is no solution to resolve the conflicts, the data could be lost or corrupted \cite{kleppmann2017designing}. This can happen during a \textit{network partition} where a client is isolated to a minority of nodes that include the leader node. If the partition recovers in a very short time, the cluster will be normal. However, if the partition lasts enough time for the majority side nodes to judge that the leader node has died and vote for another node on the majority side of the partition to be the new leader, the writes that the client has sent to the old leader on the minority side would be lost \cite{Redis2025Scale}, and hence \textit{consistency} is not guaranteed.

\begin{figure}[H]
    \centering
    \includegraphics[width=1.2\textwidth]{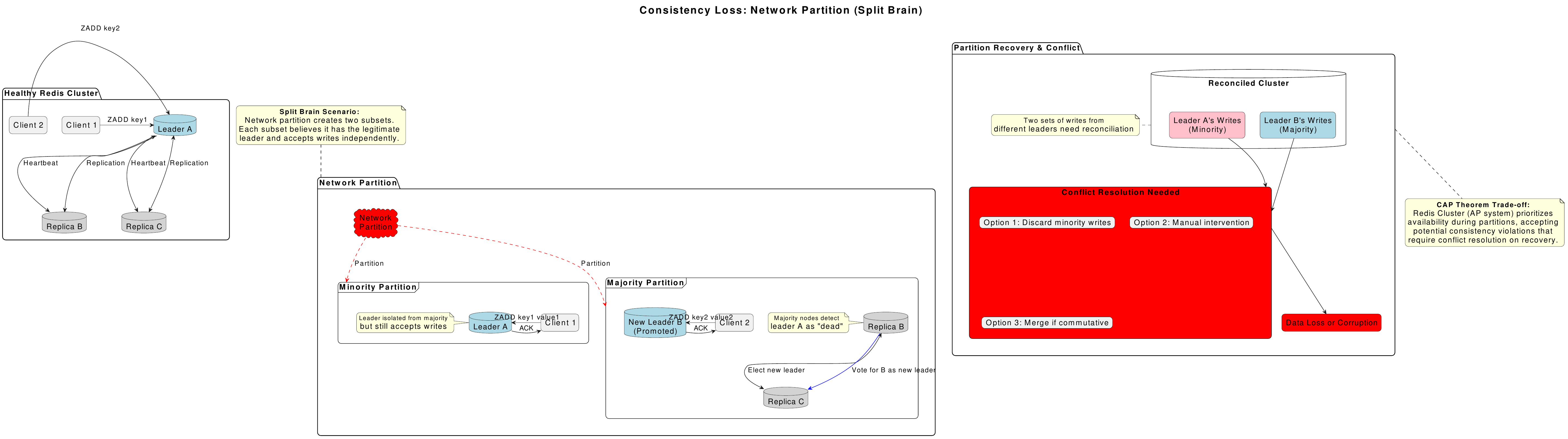}
    \caption{\textbf{Write conflict scenario during network partition in Redis Cluster (which causes split brain).} A network partition isolates the original leader from the majority of nodes. The minority partition continues accepting writes while the majority elects a new leader and also accepts writes. When the partition heals, conflicting writes from both leaders must be reconciled, potentially resulting in data loss or corruption. This demonstrates the CAP theorem trade-off where AP systems prioritize availability during partitions.}
    \label{fig:network-partition}
\end{figure}

In this design case, we explicitly favor the \textit{AP} (\textit{availability} and \textit{partition tolerance}) model over \textit{CP} (\textit{consistency} and \textit{partition tolerance}) as dictated by the CAP theorem. The scenarios illustrated in Figures \ref{fig:node-failure} and \ref{fig:network-partition} demonstrate that Redis Cluster, like many production distributed systems, cannot guarantee both strong consistency and availability during network partitions. However, our choice of AP is not merely a recognition of Redis Cluster's technical limitations but a deliberate engineering decision based on the specific requirements of rate limiting.

While the potential loss of writes in edge cases is an important consideration, the primary rationale for selecting AP stems from the utility function of rate limiting systems: \textbf{losing a few rate limit counts during a partition is preferable to dropping all legitimate traffic}. A rate limiter that becomes completely unavailable during network partitions would fail its core purpose of protecting backend resources while maintaining service accessibility. The temporary inconsistency introduced by the AP model—where different nodes might temporarily count requests differently—is an acceptable trade-off for ensuring that the rate limiting service remains operational and continues to provide basic protection even during partial system failures.

This pragmatic approach aligns with real-world production requirements where momentary rate counting inaccuracies have far less severe consequences than service-wide unavailability. By accepting eventual consistency through Redis Cluster's AP model, we prioritize system resilience and continuous operation, which are critical for maintaining API availability and user experience during the network partitions that inevitably occur in distributed deployments.

This trade-off reflects a broader design spectrum in production rate limiting systems. Our approach using Redis Sorted Sets and server-side Lua scripting provides strong atomicity for each individual rate limit decision within a single Redis node, even though the overall system pursues an AP model across the cluster. This differs from architectures optimized for extreme scale that sacrifice per-decision accuracy for throughput, such as Databricks' batch-reporting system, which accepts temporary over-limit conditions in exchange for 10x tail latency improvements \cite{Hoa2025High}. Both approaches are valid; the choice depends on whether strict per-request atomicity or maximum horizontal scalability is the primary constraint.

\subsection{Multi-Data Center Deployment}
In order to overcome high latency for clients far from the primary data center, edge computing with geographically distributed rate limiters can be deployed. In order to balance the availability and consistency trade-off, strategies for implementation can be:
\begin{itemize}
    \item Regional Redis clusters with cross-DC replication; DNS-based round robin routing to nearest rate limiter, for both availability and latency concern.
    \item An isolated counting system can be created within each Point-of-Presence (PoP), since traffic from a single IP address will always reach the same PoP under normal conditions \cite{Desgats2017How}.
\end{itemize}


\subsection{Further Optimizations}
There are certainly more optimizations, e.g., for reducing latency:
\begin{itemize}
    \item Maintain persistent connections, e.g., TCP connection pooling, by setting up Redis's connection ports to reduce connection overhead between Redis and clients \cite{Redis2025Connection}.
    \item Reduce round-trip times by batching multiple Redis commands for counter updates by the Redis pipelining technique \cite{Redis2025Serialization, Redis2025Pipelining, Hoa2025High}.
\end{itemize}

\section{Production Considerations}

\subsection{Monitoring and Analytics}
After deploying the rate limiter for a period of time, the follow-up step is to get feedback resulting from the current rate limiter's parameter settings by monitoring some infrastructure metrics, such as:
\begin{itemize}
    \item Actual \textit{worker utilization} on the server machine: monitor how the system is under pressure after applying the rate limiter \cite{Tarjan2017Scaling_blog}.
    \item Rate limit utilization: monitor request patterns, violation rates, and user behavior to optimize rate limiting policies and detect abuse patterns.
    \item Redis performance: track memory usage, latency, and throughput of the underlying Redis infrastructure to ensure optimal rate limiter performance.
    \item Anomaly detection and alerting: implement real-time detection of unusual traffic patterns, system degradation, and potential security threats with automated alerting.
\end{itemize}
As not mentioned so far, the \textit{worker utilization load shedder} is a special rate limiter, which monitors how heavy the system is under pressure and determines if accepting new requests or not accordingly, based on an active \textit{worker utilization} metric (a value between 0 and 1, evaluated depending on different application infrastructures) and tuned threshold values, rather than enforcing static business rules. It dynamically adapts to the state of the system through probabilistic shedding and slowly takes action, which is one type of \textit{flow control} mechanism. For readers interested in the mechanism and implementation of the \textit{worker utilization load shedder} rate limiter, please refer to \cite{Tarjan2017Scaling}.

\subsection{Fault Tolerance Strategies} \label{fault_toler}
In order to prevent single point of failure, \textit{Redis Sentinel} provides high availability for Redis with a distributed system that automatically detects failures of the leader nodes and replaces them with a replica node. The other additional replicas are reconfigured to use the new leader, and the clients using Redis are informed about the new address to use when connecting \cite{Redis2025High,Kong2023ZRedis}. 

For more details on advanced monitoring and alerting for scalable distributed system, and fault tolerance system design considering node failures, unreliable networks, and trade-offs between availability, consistency, durability, and latency, which we do not have space to discuss in detail in this article, keep up to date with our follow-up articles.

\section{Conclusion}
This article presents a concrete architectural design to build a scalable, distributed rate limiting system. By evaluating fundamental algorithms and analyzing their trade-offs, we established the Rolling Window model as the foundation for accurate request enforcement. Our primary contribution is the detailed design and justification of an implementation using Redis Sorted Sets and server-side Lua scripting, which together provide $O(log(N))$ operational efficiency and guaranteed atomicity to prevent race conditions.

We make the explicit architectural decision to embrace Availability and Partition Tolerance (AP) from the CAP theorem when deploying on a Redis Cluster. We argue that for a rate limiter, where global system uptime is paramount, accepting ephemeral state inconsistency during failovers is a correct and pragmatic engineering trade-off. This design choice, coupled with the inherent performance of an in-memory data store, enables the system to meet the core requirements of operational resilience, low latency, and high throughput.

The key insights from this design are: (1) algorithm selection dictates fundamental system behavior, with the Rolling Window providing accuracy at a predictable memory cost; (2) atomic operations, achieved here through Lua scripting, are non-negotiable for correctness in concurrent, distributed environments; (3) system observability and capacity planning based on memory models (e.g., ~8 bytes per request) are essential for production scalability; and (4) rules configuration via a three-layer architecture which loads the cached scripts by hashing with rule parameters without the necessary modification of the scripts, which is flexible to adapt to varying throttle rules.

Future work could explore hybrid models combining algorithm strengths, automated limit tuning based on real-time load \cite{li2023noah}, and the application of this architectural pattern to other stateful coordination problems in distributed systems. 


\section*{Acknowledgments}
This research was conducted as part of technical development at WynerTech Solutions LLC, focusing on scalable system architecture patterns.

The authors thank the Redis open-source community for developing and maintaining the in-memory data store that serves as the foundation for our implementation. We acknowledge the seminal work of Hayes \cite{Hayes2015Better} and Tarjan \cite{Tarjan2017Scaling} for their foundational work on Redis Sorted Set rate limiting.

\bibliographystyle{IEEEtran}
\bibliography{ref}

\end{document}